\documentstyle[prl,aps,epsf,twocolumn]{revtex}

\tighten

\begin{document}
\draft
\twocolumn[\hsize\textwidth\columnwidth\hsize\csname @twocolumnfalse\endcsname

\title{\bf High-quality continuous random networks}

\author{G.T. Barkema}
\address{
Theoretical Physics, Utrecht University, Princetonplein 5, 3584 CC
Utrecht, the Netherlands \protect\cite{barkadd} }

\author{Normand Mousseau}

\address{
Department of Physics and Astronomy and CMSS, Ohio University, Athens, OH
45701, USA \protect\cite{mousadd}}

\date{Submitted to the Physical Review B - \today}

\maketitle


\begin{abstract}

The continuous random network (CRN) model is an idealized model for
perfectly coordinated amorphous semiconductors.  The quality of a CRN
can be assessed in terms of topological and configurational properties,
including coordination, bond-angle distributions and deformation
energy.  Using a variation on the {\it sillium} approach proposed 14
years ago by Wooten, Winer and Weaire, we present 1000-atom and
4096-atom configurations with a degree of strain significantly less
than the best CRN available at the moment and, for the first time,
comparable to experimental results.

\end{abstract}

]

\vspace*{-0.5cm} \narrowtext

\section{Introduction}

The structure of amorphous semiconductors, as seen by theorists, is
well represented by the continuous random network (CRN) model
introduced more than 60 years ago by Zachariasen \cite{zachariasen32}.
The interest of this model lies in its simplicity: the only requirement
of this model is that each atom should satisfy fully its bonding needs.
In addition, the quality of a CRN is generally determined by the amount
of strain, as measured by the local deviations from crystalline
environment, induced by this constraint; the ``ideal'' CRN being
typically defined as that with the lowest spread in the bond length and
bond angle distributions.

In spite of the simplicity of the model, it has turned out to be
difficult to actually prepare CRN realizations of a quality comparable
to that of experiment, making it difficult to fully assess the real
structure of amorphous semiconductors.  The origin of this problem has
generally been attributed to weaknesses in the model-building
community: standard approaches such as molecular dynamics cannot reach
time scales appropriate for full relaxation.  Moreover, other
techniques suggest that empirical and semi-empirical potentials able to
reproduce all properties of amorphous semiconductors are still
missing\cite{ding,cook93,art}. An alternative explanation for the fact
that experimentally a lower spread in bond length and angular
distribution is observed, might be that the coordination in
high-quality {\it a}-Si samples is significantly lower than four.
Laaziri et al. report a coordination as low as 3.88 \cite{laaziri99}, a
density of defects orders of magnitude higher than what is measured
using electron-spin resonance techniques \cite{laaziri99}.  If true,
this higher density of defects might easily facilitate a lower spread
in the bond lengths and angles, explaining in part the discrepancy
between experiment and theoretical models.

Following a long tradition, one approach to shed some light on this
discrepancy is to try to see how far it is possible to push the
continuous random model in order to reach structural properties in
agreement with experiment. By creating idealized networks with the same
angular deviation as the experimental ones and a good overall fit to
the radial distribution function, it is possible to show that perfect
coordination in amorphous silicon is not ruled out by the low angular
deviation.  This is the purpose of this paper  which follows a long
series of works going in the same direction
\cite{guttman80,www,holender91,djordjevic95}.

Using a modified version of the Wooten-Winer-Weaire algorithm, we have
succeeded in creating a number of totally independent 1000-atom
configurations with a bond-angle distribution as low as 9.19 degrees,
almost 2 degrees below the best available numerical models without
four-membered rings and on a par with experimental
values~\cite{khalid99}. The algorithm we use avoids completely the
crystalline state, contrary to previous WWW-type approaches.  Moreover,
as shown below, the structural and electronic properties of the
networks are excellent, making them ideal starting point for empirical
as well as tight-binding or {\it ab-initio} studies~\cite{lewis98}.

This paper is constructed as follows. First, we review briefly the
Wooten-Winer-Weaire algorithm and detail our simulation procedure.
Next, we present structural and electronic properties of the
configurations generated, and compare them with previous simulations and
experimental results.

\section{Methodology and Details of Simulations}

\begin{figure}
\begin{center}
\epsfxsize=6cm
\epsfbox{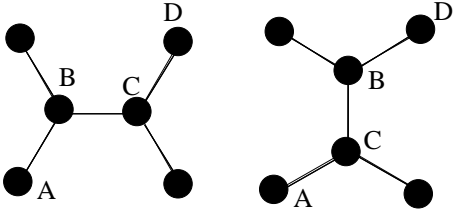}
\end{center}
\caption{A basic WWW event. Left: before; right: after the bond
exchange.}
\label{fig:www}
\end{figure}

In the {\it sillium} approach\cite{www}, proposed by Wooten, Winer and Weaire
(WWW) to generate CRN structures, a configuration consists of the
coordinates of all $N$ atoms, together with a list of the $2N$ bonds
between them.  The structural evolution consists of a sequence of bond
transpositions involving four atoms.  Four atoms A, B, C, and D are
selected following the geometry shown in Fig.  \ref{fig:www}; two
bonds, AB and CD, are then broken, and atoms A and D reassigned,
respectively, to C and B, creating two new bonds, AC and BD. After the
transposition, all atoms are allowed to relax within the constraints of
the neighbor list.

Within this approach, the generation of a CRN starts with a cubic diamond
structure, which is then randomized by a large number of bond
transpositions.  After  thermalization, the network is relaxed through a
sequence of many more proposed bond transpositions, accepted with the
Metropolis acceptance probability
\begin{equation}
P={\rm Min}\left[ 1, \exp((E_b-E_f)/k_bT) \right],
\label{Metrop}
\end{equation}
where $k_b$ is the Boltzmann constant, $T$ is the temperature, and $E_b$
and $E_f$ are the total energies of the system before and after the
proposed bond transposition.

The list of neighbors determines the topology, but also the energy of
the network: independently of the distance between two atoms, they
interact only if they are connected in the list of neighbors. With an
explicit list of neighbors, it is possible to use a simple interaction
such as the Keating potential~\cite{keating66}:
\begin{eqnarray}
E&=&\frac{3}{16} \frac{\alpha}{d^2}
\sum_{\langle ij \rangle}\left( {\bf r}_{ij} \cdot {\bf r}_{ij}-d^2\right)^2
\nonumber \\
& &+\frac{3}{8} \frac{\beta}{d^2} 
\sum_{\langle jik \rangle} \left( {\bf r}_{ij} \cdot {\bf r}_{ik}+\frac{1}{3}d^2
\right)^2,
\end{eqnarray}
where $\alpha$ and $\beta$ are the bond-stretching and bond-bending
force constants, and $d=2.35$ \AA\, is the Si-Si strain-free equilibrium
bond length in the diamond structure.  Usual values are $\alpha=2.965
$eV/\AA$^2$ and $\beta=0.285\; \alpha$. 

With the approach described above, along with a few more details that
can be found in Ref. \cite{www}, Wooten and Weaire obtained 216-atom
structures with an angular distribution as low as 10.9 degrees. A decade
later, using the same approach but more computing power, Djordjevi\'c,
Thorpe and Wooten (DTW) produced some large (4096-atom) networks of even
better quality, with a bond angle distribution  of 11.02 degrees for
configurations without four-membered rings and 10.51 degrees when
these rings where allowed~\cite{djordjevic95}. 

In the present work, using a series of algorithmic improvements and
faster computers, we are able to generate structurally and
electronically better networks:  the 1000-atom configurations, for
example, show a bond-angle distribution of almost two degrees lower than
DTW's model while our 4096-atom cell is more than one degree better.

The improvements introduced to the sillium approach are the following:
\begin{enumerate}
\item  we start from a truly random configuration rather than from a
molten crystalline state, thus guaranteeing that the structure is not
contaminated by some memory of the crystalline state;

\item  we evaluate the acceptance of a trial move using a Metropolis 
accept/reject procedure {\it without} doing full relaxation;

\item  we use a local/non-local relaxation procedure to limit the number of
force evaluations, i.e., we relax only locally in the first ten relaxation
steps after a bond transposition (up to the third neighbor shell); in
combination with 2), this makes the time per bond transposition almost
independent of the configuration size;

\item  at regular times, we quench the structure at zero temperature, with
advantages outlined in section \ref{quench}.
\end{enumerate}

With these improvements, the generation of the networks goes as
follows. We first generate starting configurations as described in
section \ref{starting}, and quench these structures as described
in section \ref{quench}.  Next, we alternate running at a temperature
of 0.25 eV for about 100 trial bond transpositions per atom, and
quenching.  The decrease in energy is almost exclusively obtained
during the quenching, the role of the annealing at finite temperature
is mostly to provide for a fresh starting point for the next quench.
Once the energy is brought down to about 0.3 eV per atom, and the
angular spread around 10 degrees, this procedure yields
diminishing returns: the annealing is no longer able to bring the
sample to a starting point where the quenching leads to a lower
minimum. To lower the energy further, we therefore also anneal the
configurations in different conditions for a few hundred trial bond
transpositions per atom, like a stronger three-body force or a larger
volume.

\subsection{Generating random initial CRNs}
\label{starting}

\begin{figure}
\begin{center}
\epsfxsize=6cm
\epsfbox{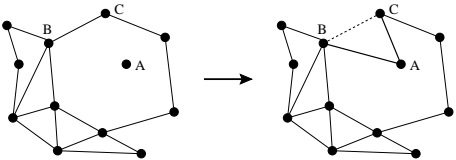}
\end{center}

\vspace{0.5cm}
\caption{One step in the expansion of the loop, that eventually visits
all atoms four times. Three atoms A, B, and C are involved, of which B
and C are bonded, while A is bonded to neither B nor C, and of which A
is not four-fold coordinated. The bond BC is then replaced by bonds AB
and AC.}
\label{fig:loopexp}
\end{figure}

To generate a random initial configuration, we randomly place the atoms
in a box at crystalline density, under the constraint that no two atoms
be closer than 2.3 \AA. The difficult part is to connect these atoms in
order to obtain a tetravalent network. We achieve this by starting
with a loop visiting four atoms somewhere in the sample, in such a way
that each pair of atoms that are neighbors along the loop be not
separated by more than a cut-off distance $r_c$. This loop is gradually
expanded until it visits each atom exactly twice; the steps of the loop
are then the bonds in our tetravalent network. The expansion of the loop
is achieved by randomly selecting a group of three atoms A, B and C,
such that A is not four-fold coordinated and is within a distance of
$r_c$ from B and C but not bonded to either, while B and C are bonded.
Next, the bond BC is replaced by bonds AB and AC, expanding the loop by
one step. This procedure is illustrated in figure \ref{fig:loopexp}.
Initially, $r_c$ is set to some small value like 3 \AA, but then it is
gradually increased until all atoms are four-fold coordinated.  Although
this method leads to highly strained initial configurations, it has the
advantage that it contains absolutely no trace of crystallinity.

This process leads typically to CRNs whose angular distribution
initially has a width of around thirty degrees, but which reduces
rapidly to around 13 degrees in a single quench (see section
\ref{quench}). In the beginning of this first quench, when the angular
deviation is quite large, sometimes a pair of atoms is closeby without
being bonded; to eliminate such artefacts that result from the fact
that within the Keating potential atoms only interact if they are
explicitly bonded, we replace a bond of each of these atoms by a bond
between these atoms and another bond between their neighbors (conserving
four-fold coordination).

\subsection{Avoiding complete relaxation of trial configurations}

In the standard sillium approach, a move consists of a bond
transposition followed by full structural relaxation and an
accept/reject step according to the Metropolis criterion
(\ref{Metrop}).  An alternative implementation is to first decide a
threshold energy given by
\begin{equation}
E_t=E_b-k_bT \log(s),
\label{threshold}
\end{equation}
where $s$ is a random number between 0 and 1. The proposed move is then
accepted only if $E_f \leq E_t$. This procedure is exactly equivalent to
the usual Metropolis procedure. By fixing the threshold for $E_f$ {\it
before} a transposition is attempted, it is however possible to reject
the move as soon as it becomes clear that this threshold cannot be
reached, i.e., before the configuration is fully relaxed.  Since the
energy is harmonic around the minimum, the decrease in energy obtained
during further relaxation is approximately equal to the square of the
force times some proportionality constant $c_f$, so that during the
relaxation the final energy can be estimated to be
\begin{equation} E_f \approx E-c_f |F|^2.  \end{equation}
If, at any moment during the relaxation, $E-c_f |F|^2 >E_t$, the trial
move is rejected and a new one is started.  Such a method requires, of
course, a conservative choice for $c_f$; in our units, the
proportionality constant $c_f$ in well-relaxed configurations is always well
below 1. To account for anharmonicities, we do not reject any move
during the first five steps of relaxation.

Since much less than one percent of the proposed moves are accepted in
well-relaxed samples, avoiding spending time on moves that are
eventually rejected can produce a significant gain in efficiency; using
this improvement, we observed a speed-up of close to an order of
magnitude, so that all bond transpositions in a 1000-atom network can be
attempted in less then 3 minutes on a 500 Mhz DEC-Alpha workstation. 

\subsection{Efficient quenching}
\label{quench}

Further optimizations are possible in the case of zero temperature.
Since the threshold energy (Eq. (\ref{threshold})) is constant, a 
proposed bond transposition that is once rejected will keep on being
rejected, as long as no other bond transpositions are accepted in
the mean time. A combination of four atoms ABCD with bonds AB, BC and CD
can be selected in $N\cdot 4\cdot 3\cdot 3/2$ times, so there are
$18N$ possible bond transpositions. We mark all bond
transpositions that were rejected since the last accepted bond
transposition to avoid retrying these. Once all bond transpositions
have been tried but rejected, the quenching is complete. At this stage,
the system is not only at a local energy minimum (i.e. a point in
phase space where the force is zero and all eigenvalues of the hessian
are positive), but no single bond transposition can lower the
energy. The configurations we discuss here, have this property.

In the standard sillium approach, the creation of four-membered rings is
disallowed. Following DTW \cite{djordjevic95}, we find that especially
for quenching the relaxation is significantly helped by allowing for
four-membered rings, because of the large extra number of pathways
accessible to the system. At the end of the quenching, the few four-membered
rings that are created can easily be removed one by one, by chosing the
energetically most favorable bond transposition in which bond AB is
part of the four-membered ring (and where no new four-membered rings are
introduced).  Typically, the energy increases by less than an eV per
removed four-membered ring.

\section{Resulting configurations}

We present here results for three different configurations: two
1000-atom cells and one with 4096 atoms.  In Table \ref{tab:kea}, we
compare our configurations, relaxed with the Keating potential used in
the modified WWW algorithm, with those of Djordjevi\'c, Thorpe and
Wooten \cite{djordjevic95}.  We also provide the irreducible ring
statistics, i.e., those rings in which no two atoms are connected
through a sequence of bonds that is shorter than the sequence along the
ring.  We also provide the ring statistics for all $n$-membered
rings in order to compare with Ref. \cite{www} and other papers in the
literature (Table \ref{tab:msw}).
Table \ref{tab:kea} shows that the strain per atom in our structures is
significantly below that of DTW.  One of the standard measurements to
evaluate the quality of a model is the coordination number as computed
based on the radial distribution function (RDF).  Using the minimum of
the RDF between the first- and second--neighbor peak and after
relaxation with the Keating model, the first two configurations are
perfectly tetravalent. The 4096-atom configuration has 0.1\% of 5-fold
defects.

\begin{table}
\caption{Energetic and structural properties of models relaxed with the
Keating potential. The first two models, DTW$^{(1)}$ and DTW$^{(2)}$
are the models prepared in \protect\cite{djordjevic95} and refer,
respectively, to a model with and without four-membered rings.
Configurations 2 and 3 are 1000-atom models prepared according to the
procedure described here and `4096' is a 4096-atom model prepared the
same way. All three models are without four--membered rings.  The ring
statistics are for irreducible rings and $\rho_0$ is based on $r_0=2.35$ \AA.}

\label{tab:kea}
\begin{tabular}{c|ccc|cc|c}
                         & DTW$^{(1)}$ & DTW$^{(2)}$  &Conf. 2&Conf. 3& 4096 
\\ \hline 
E(eV)/atom               & 0.336  & 0.367  & 0.267  &  0.264 & 0.304\\
$\rho/\rho_0$            & 1.000  & 1.000  & 1.043  & 1.040  & 1.051 \\
$\langle r \rangle/r_0$  & 0.996  & 0.997  & 0.982  & 0.982  & 0.980 \\
$\Delta r/r_0 $ (\%)     & 2.52   & 2.65   & 3.94   & 0.371  & 4.17 \\
$\langle \theta \rangle$ & 109.24 & 109.25 & 109.30 & 109.27 & 109.28\\
$\Delta \theta $         & 10.51  & 11.02  & 9.21   &  9.20  & 9.89 \\
   \\
Rings/atom \\
4                        & 0.015  & 0.000  & 0.000  & 0.000 & 0.000\\
5                        & 0.491  & 0.523  & 0.472  & 0.480 & 0.490\\
6                        & 0.698  & 0.676  & 0.761  & 0.750 & 0.739\\
7                        & 0.484  & 0.462  & 0.507  & 0.515 & 0.467\\
8                        & 0.156  & 0.164  & 0.125  & 0.116 & 0.148\\
9                        &        &        & 0.034  & 0.033 & 0.035\\
\end{tabular}
\end{table}

Another important quantity that can be compared with experiment is the
width of the bond angle distribution.  Experimentally this quantity can
be extracted from the radial distribution function, or from the Raman
spectrum --- using a relation proposed by Beeman {\it et
al.}\cite{beeman}.  The most recent measurement, taken on annealed
samples prepared by ion bombardment and using the second-neighbor peak
of the radial distribution function, gives 10.45 and 9.63 degrees,
respectively, for as-implanted and annealed samples~\cite{khalid99}.
Our configurations, relaxed with the Keating potential, present
therefore a bond angle distribution slightly narrower than experimental
values.  (This is to be expected of the ``right'' structure since the
theoretical models are taken at zero K.)

\begin{figure}
\epsfxsize=8cm
\epsfbox{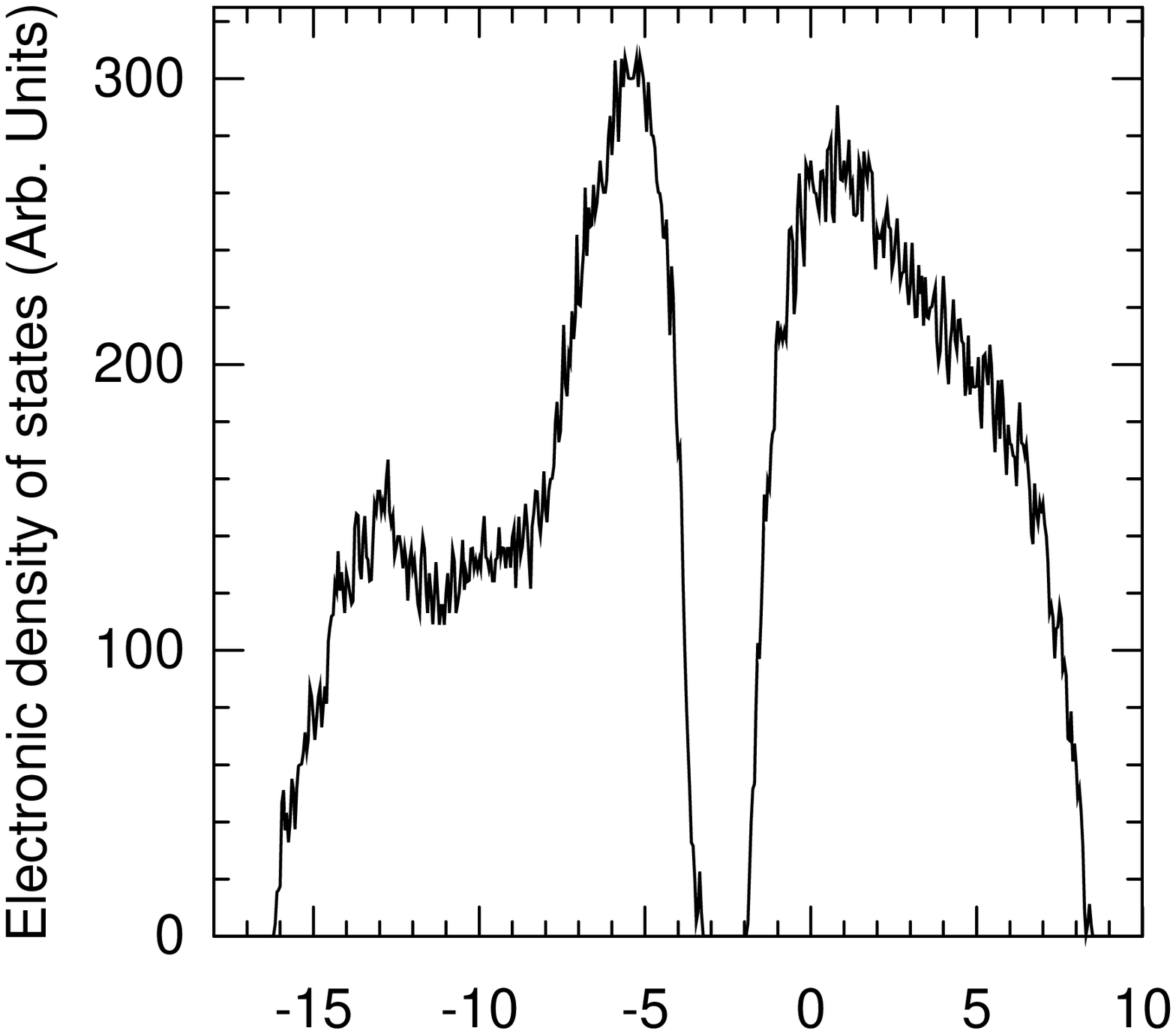}

\epsfxsize=8cm
\epsfbox{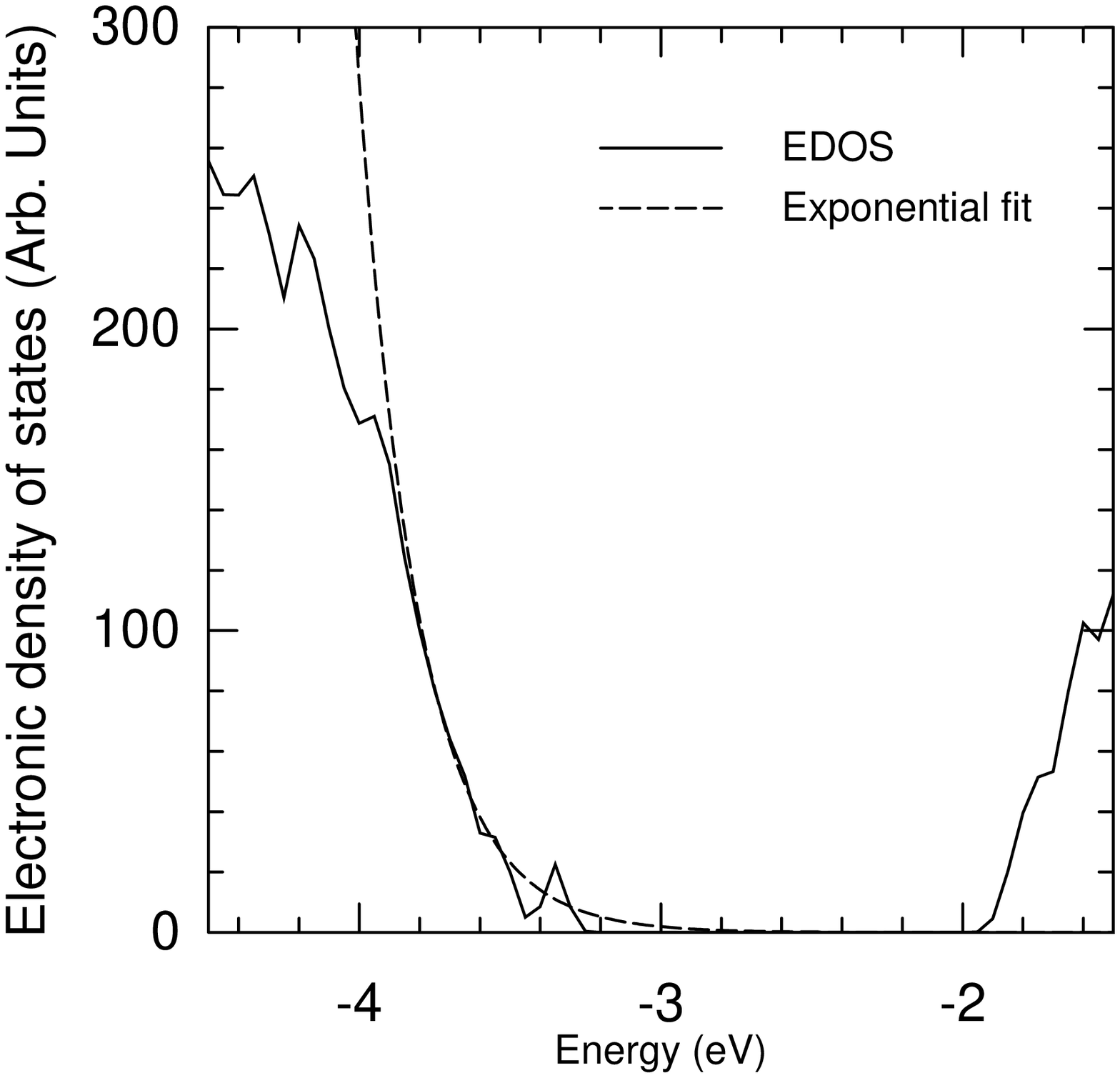}

\caption{ Top:
Electronic density of states for the 1000-atom model 2 as obtained using
ab-initio tight-binding. Bottom: Close-up on the gap region. The
dashed curve is an exponential fit, $\rho(E) \propto \exp(-E/E_0)$, with
$E_0=0.2 eV$.
\label{fig:edos}}
\end{figure}

While structural averages provide a good idea of the overall quality of
a model, they do not say much regarding the local environments. It is
therefore important to look also at the electronic properties of these
models: even small densities of highly strained geometries or defect
atoms will be picked up as states in the gap of the electronic density
of states (EDOS). In the last few years, it has become possible to
compute the electronic structure of multi-thousand-atom
configurations.
Here, we show the electronic density of state for
Configuration 2, a 1000-atom configuration. Because of the costs of
doing a full ab-initio atomic relaxation, we have relaxed the
cell with the Keating potential and used the Fireball
local-basis ab-initio code to obtain the electronic density of
state\cite{fireball}.  Previous work showed that configurations relaxed
with a Keating potential demonstrated little further relaxation with the
Fireball code~\cite{dong98} so that the results presented here are
unlikely to change very much during further relaxation.  Figure
\ref{fig:edos} shows the EDOS smoothed with a gaussian of width 0.01 eV.
A remarkable feature of this EDOS is the absence of states in the gap,
leading to a perfect gap of 1.3 eV.  The generation of defect-less
models is very important for our understanding of the electronic
dynamics and the role of defects in disordered semiconductors.  The
decay of the valence tail, the Urbach tails, can be reasonably well
approximated by an exponential --- $\rho(E) \propto \exp(-E/E_0)$ --- with
$E_0 = 0.2$ eV, in agreement with previous calculations~\cite{dong98}.
                                            
Although we get good structures using the Keating potential, it is
important to verify the stability of these networks when relaxed with a
more realistic interaction potential that does not require a pre-set
list of neighbors. There exists no empirical potential at the
moment that can describe fully the properties of {\it a}-Si; we use a
modified Stillinger-Weber potential, where the three-body contribution
to the energy is enhanced by 50 \% with respect to the two-body term.
This ad-hoc modification was shown to produce good structural properties for
amorphous silicon~\cite{ding,holender91,art,art99}. 

After relaxation at zero pressure, the two 1000-atom configurations
remain perfectly coordinated. The 4096-atom cell, less well relaxed,
develops a few coordination defects : based on the first minimum in the
RDF,  it presents respectively 0.4\% and 0.3\% of three-fold and five-fold
coordinated atoms.

Table \ref{tab:msw} presents the structural and energetic properties of
the relaxed CRNs at zero pressure. For all configurations, the bond
angle distribution widens and the density decreases significantly
compared to the Keating-relaxed structures.  For the 1000-atom
configurations, the local relaxation with the modified Stillinger-Weber
potential did not result in a change of topology and the total energies
are very low compared with previous models\cite{artb}.  We therefore
expect that the structures will be stable with any reasonable
potential.

\begin{figure} 
\epsfxsize=8cm 
\epsfbox{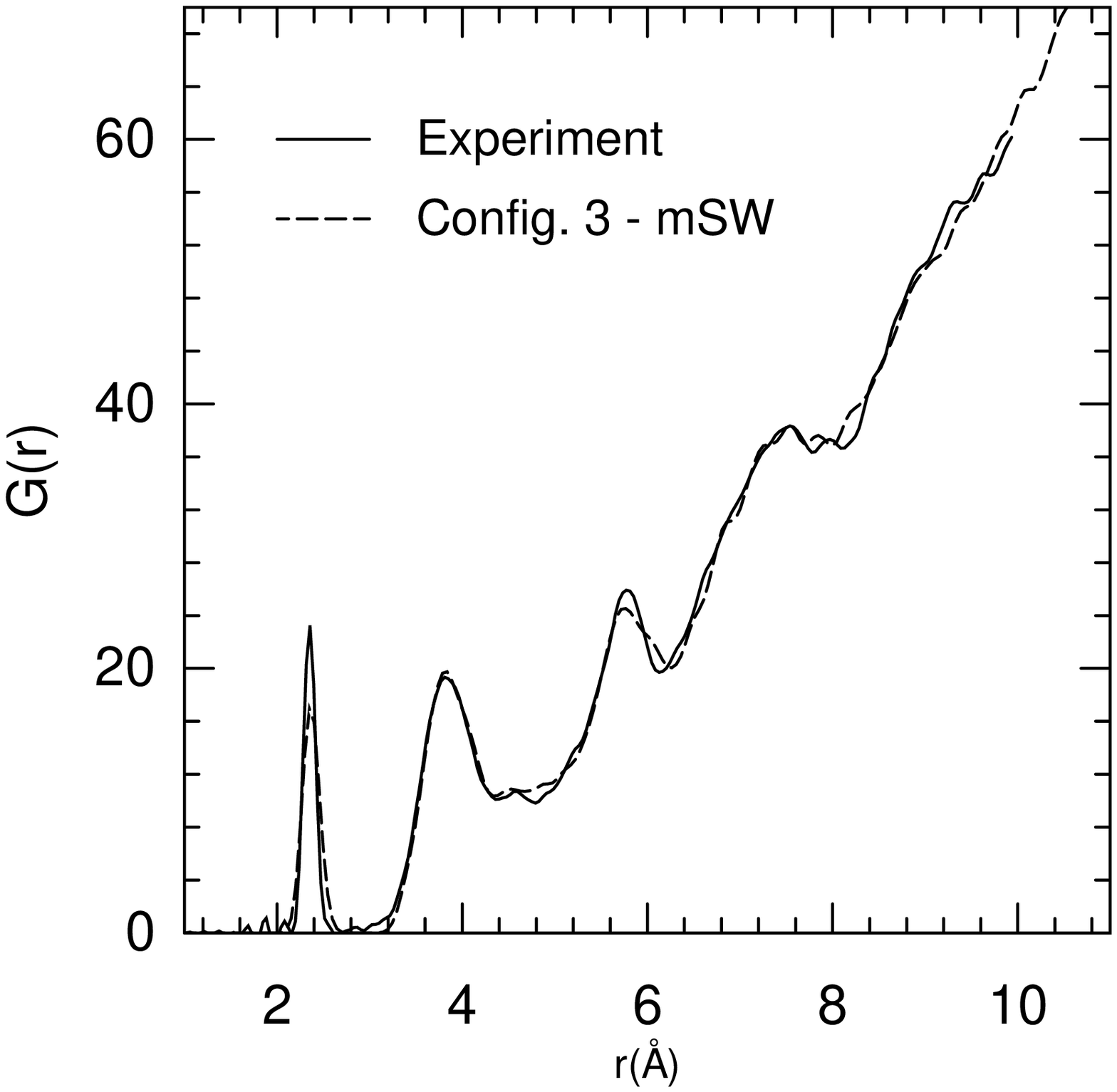}

\epsfxsize=8cm
\epsfbox{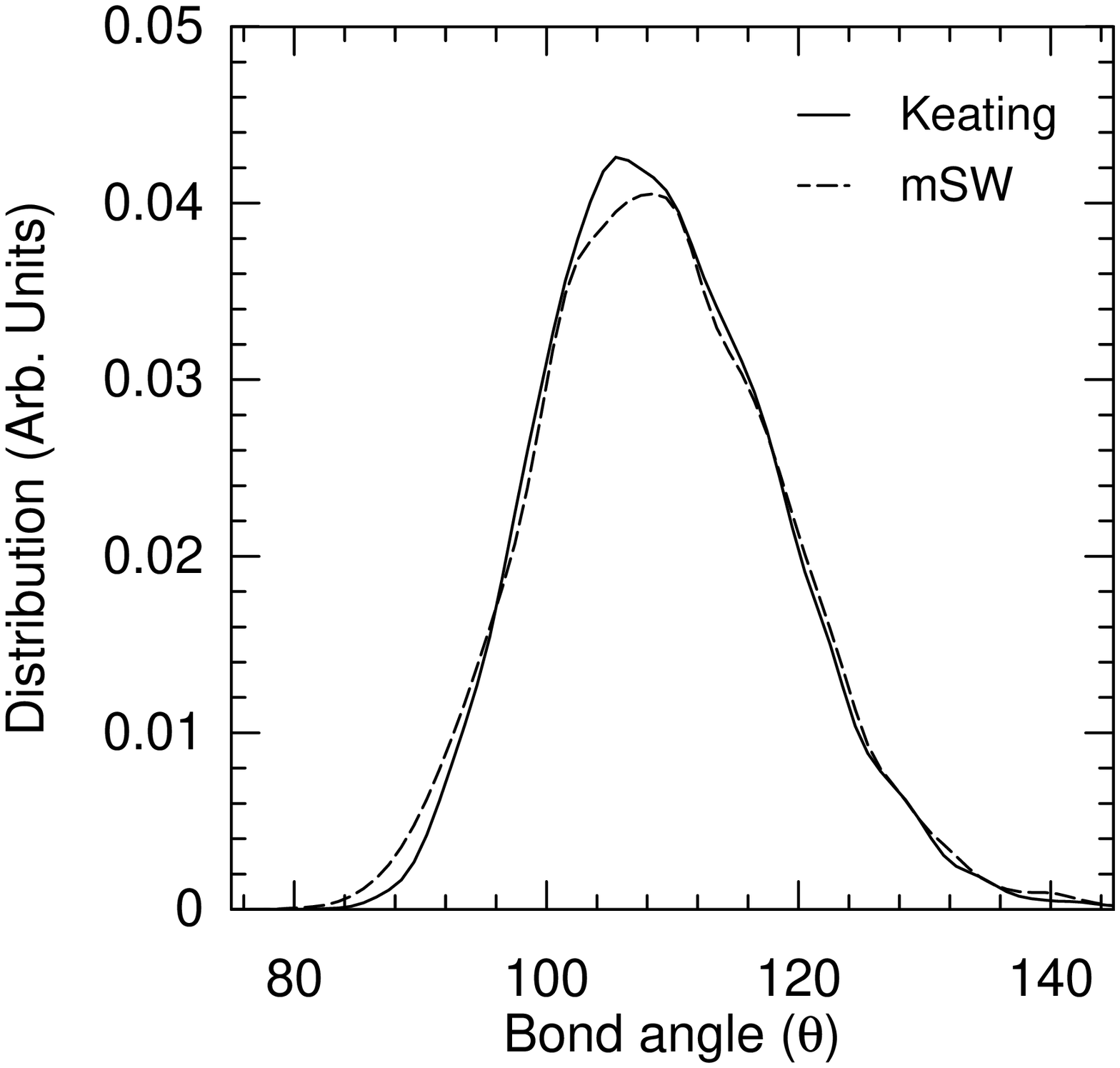}

\caption{Top: Radial distribution function for Configuration 3 relaxed
with the modified Stillinger-Weber potential.  Solid line: experimental
results from Ref. \protect\cite{khalid99}. To obtain a better fit, the
computer-generated structure is scaled by a linear factor of 0.99.
Bottom: Bond angle distribution for Configuration 3 relaxed with
Keating and the modified Stillinger-Weber potential. The curve is
smoothed with a gaussian of width 2 degrees.}
\label{fig:rdf}
\end{figure}

Figure \ref{fig:rdf} shows a comparison of Configuration 3 with
experimental data obtained by Laaziri {\it et al.} on annealed {\it
a}-Si samples prepared by ion bombardment. The agreement between the
two is excellent except for some discrepancy in the height of the
third-neighbor peak.  Such an agreement must only be seen as a sign
that the topology might be right, however: configurations differing
widely in their topology can easily produce similar radial distribution
functions ~\cite{gaas}.  The same figure also presents the bond-angle
distribution for Configuration 3 relaxed with both the Keating and the
modified Stillinger-Weber potentials.  As expected for a perfectly
coordinated sample, the distribution is smooth and presents  a single
peak centered at the tetrahedral angle.

To compare with previous molecular dynamics studies~\cite{md}, we have
also relaxed our cells with the standard Stillinger-Weber potential,
which is known to give an incorrect amorphous structure.  After
relaxation of Configuration 2, we find 17 atoms with five-fold coordination
and three three-fold coordinated ones; similar results are found with the
two other networks.  The resulting configurational energy, given in
Table \ref{tab:msw}, compares favorably with molecular
dynamical results. 

Contrary to the topological properties, which seem relatively
independent of the details of the potential, we see that the ideal
density of amorphous silicon compared with that of the crystal changes
qualitatively as a function of the potential used. Configurations
relaxed at zero pressure with the Keating potential show a reproducible
densification by  2 \% while the modified Stillinger-Weber
potential, also at zero pressure, leads to a structure which is
up to 6 \% less dense.  The latter results are in qualitative agreement with
experiment~\cite{custer94} and previous simulations using a similarly modified
potential~\cite{holender91}.  Recently, Laaziri and collaborators have
pointed to the lower density of {\it a}-Si as an explanation for the
relatively low coordination measured by x-ray scattering. Our results,
on the contrary, indicate that there is very little dependence between
the density of the amorphous material and its topology, at least within
the application of our two empirical potentials.  A volume change at the
percent level should therefore have very little impact on the topology and
will reflect mostly some fine details of the real atomic interactions. 
             
\begin{table}

\caption{Structural properties of our configurations after relaxation
with the modified Stillinger-Weber (mSW) potential. Except for the
4096-atom configuration, the topology has remained unchanged (see text.)
For comparison purposes, the total ring number per atom (including
reducible ones) is also reported, as well as the energy after relaxation
with the original Stillinger-Weber (SW) potential; MD-prepared samples
give -4.088 eV/atom \protect\cite{lewis95}.  }

\label{tab:msw}
\begin{tabular}{c|ccc|cc|c}
                         &Sample 2&Sample 3& 4096 \\ \hline 
E(eV)/atom (mSW)         & -4.026 & -4.034 & -3.990 \\
E(eV)/atom (SW)          & -4.126 & -4.133 & -4.106 \\
$\rho/\rho_0$            & 0.947  & 0.950  & 0.936\\
$\langle r \rangle/r_0$                & 1.018  & 1.017  & 1.020\\
$\Delta r/r_0 $ (\%)     & 2.9    & 2.7    & 0.032 \\
$\langle \theta \rangle$ & 109.25 & 109.24 & 109.20  \\
$\Delta \theta $         & 9.77   & 9.70  &  10.51 \\
   \\
Rings/atom \\
4                        & 0.000  & 0.000  & 0.001 \\
5                        & 0.472  & 0.480  & 0.489 \\
6                        & 0.840  & 0.847  & 0.830 \\
7                        & 1.011  & 1.023  & 0.979 \\
8                        & 2.025  & 2.002  & 2.064 \\
\end{tabular}
\end{table}

\section{Conclusions}

We have presented here modifications to the Wooten-Winer-Weaire
algorithm that allows one to produce large high-quality continuous
random networks without passing at all by a crystalline phase.
Structural and electronic properties of the networks produced are
excellent and they compare favorably with experiment. 

The coordinates of the three configurations discussed here, as well as a
10~000-atom sample under preparation, are available by request.

\section{Acknowledgements}   

We thank S. Roorda for providing us with his experimental data, and S.
Nakhmanson for computing the electronic density of states of
Configuration 2. NM acknowledges stimulating discussion with D. A.
Drabold and partial support by the National Science Foundation under
grant number DMR-9805848.

\bibliographystyle{prsty}

\end{document}